\documentstyle[12pt]{article}
\begin{document}
\def\singlespacing{\baselineskip=12pt}
\def\doublespacing{\baselineskip=24pt}
\doublespacing                      

\title{Cosmological Time in Quantum Supergravity}

\author{Robert Graham\thanks{e-mail address: graham@uni-essen.de}\\
Fachbereich Physik,\\ 
Universit\"at Gesmthochschule Essen,\\
D-45117 Essen,\\ 
Germany\\ 
{}\\
\and 
Hugh Luckock\thanks{e-mail address: luckock\underline{ }h@maths.su.oz.au}\\
School of Mathematics and Statistics,\\
University of Sydney,\\
NSW 2006,\\ 
Australia}  
\date{March 28, 1996}   

\maketitle
\begin{abstract}           
The version of supergravity formulated by Ogievetsky and Sokatchev is almost 
identical to the conventional $N=1$ theory, except that the cosmological 
constant $\Lambda$ appears as a dynamical variable which is constant only by 
virtue of the field equations. 
We consider the canonical quantisation of this theory, and show that the 
wave function evolves with respect to a dynamical variable which can be 
interpreted as a cosmological time parameter. The square of the modulus of 
the wave function obeys a set of simple conservation equations and can be 
interpreted as a probability density functional. The usual problems associated 
with time in quantum gravity are avoided.   
\end{abstract}

\leftline{PACS numbers: 98.80.Hw, 04.65.+e}
\newpage
\pagestyle{plain}  

One of the most fundamental questions in quantum cosmology is that of 
identifying a suitable time parameter, with respect to which the dynamics of 
the Universe can be  measured. The conventional Wheeler-DeWitt formulation 
gives a time-independent quantum theory, and does not suggest any obvious 
reason why observers should experience the passage of time. Moreover, the 
absence of any special parametrisation leads to ambiguities when defining 
transition 
amplitudes between specified 3-geometries \cite{Kuchar1}. A related problem is 
that of choosing an inner product on the Hilbert space of physical states, in 
the absence of any parameter with respect to which this inner product should 
be conserved.  
                   
Supersymmetry transformations are more fundamental than time translations, 
in the sense that the latter may be generated by anticommutators of 
supersymmetry 
generators. For this reason, it is natural to look to supersymmetry 
for a solution 
to the problems outlined above. One is therefore led to consider 
theories such as 
$N=1$ supergravity. 

Perhaps the most elegant and economical description of supergravity                       
is that of Ogievetsky and Sokatchev \cite{OgSok}. In this formulation, the 
supergravity multiplet is obtained from a complex vector superfield by 
imposing simple gauge conditions together with a type of unimodularity 
condition on the supercoordinate transformations. The resulting theory 
is identical in most respects to conventional supergravity (as described 
by Wess and Bagger, for example \cite{WB}).  

However there is a subtle difference between the multiplet of 
Ogievetsky and Sokatchev and that of the conventional theory, which does not 
appear to have been previously exploited \cite{AuNiTo}. The conventional 
supergravity multiplet contains a scalar density $M$ which is treated as an 
independent auxiliary field; however, in the theory of Ogievetsky and Sokatchev 
$M$ is given in terms of a vector density whose longitudinal component is 
actually dynamical. It turns out that this additional dynamical degree of 
freedom is essentially the cosmological ``constant" $\Lambda$, which in this 
case is constant only by virtue of the equations of motion.                                                
                                                                              
A dynamical $\Lambda$ also appears in general relativity if a unimodular 
condition is imposed on the metric before the variation of the fields 
\cite{Unruh,Henneaux,UnruhWald}. In that case, $\Lambda$ is found to be 
canonically conjugate to a parameter which is naturally interpreted as 
cosmological time.  

It is shown below that a similar result is obtained in the Ogievetsky
and Sokatchev version of supergravity. However in this case, unlike that 
of unimodular general relativity, the result is achieved without the 
imposition of any ad-hoc constraints on the physical fields. Moreover        
the cosmological time is a dynamical variable determined by the fields 
on a given hypersurface, while in unimodular general relativity it depends 
on the entire past history of the fields prior to this hypersurface.
  
The Lagrangian for $N=1$ supergravity is 
\begin{eqnarray} 
{\cal L} & = & -\frac12 e {\cal R}  + \frac12 e 
\epsilon^{klmn} e_{\ell}{}^a (\bar\psi_k \bar\sigma_a
{\widetilde{\cal D}}_m \psi_n - \psi_k \sigma_a
{\widetilde{\cal D}}_m \bar\psi_n)  + \frac13 e b^ab_a \nonumber \\
& & - e \lambda^*( M + \psi_a \sigma^{ab} \psi_b ) 
- e\lambda( M^* + \bar\psi_a \bar\sigma^{ab} \bar\psi_b )
- \frac13 e M^* M     
\label{eq:Lagrangian} 
\end{eqnarray} 
where $\lambda$, $\lambda^*$ are externally specfied constants, 
$e$ is the determinant of the tetrad $e_{\hat\ell} {}^a$, and 
${\cal R}$ is the Ricci scalar \footnote{We use the notation and conventions 
of \cite{WB}, with spacetime coordinates denoted by letters from the middle 
of the alphabet. Note that ${\cal R}$ differs by a sign from the 
curvature scalar defined in \cite{MTW} and \cite{D'Eath}.}. 
Conventionally, $M$ and $M^*$ are treated as auxiliary fields and eliminated 
using their equations of motion $M=-3\lambda$ and $M^*=-3\lambda^*$.
This leads to a theory with fixed gravitino 
mass $m=|\lambda|$ and cosmological constant $\Lambda=-3m^2$.

In the formulation of Ogievetsky and Sokatchev, the cosmological 
constant arises in quite a different way \cite{OgSok}. In this approach 
$M$ is not an independent field, but is given by the expression  
\begin{equation} 
M =  e^{-1} \partial_m M^m - \psi_a \sigma^{ab} \psi_b 
\label{eq:def_M}                                        
\end{equation}                                            
where the complex vector density $M^m$ now plays the part of the auxiliary 
field. The $\lambda$,$\lambda^*$ terms appearing in the Lagrangian 
(\ref{eq:Lagrangian}) are then just total divergences and can be dropped. 
Varying  $M^m$ (rather than $M$ as in the usual approach) leads to the field 
equations $\partial_m M=\partial_m M^*=0$. The equations of motion for the 
remaining fields are identical to those of conventional supergravity with 
gravitino mass $m= {1\over 3} |M|$ and cosmological constant 
$\Lambda=  -3m^2$. In the present formulation, however, $M$ is not an 
externally specified constant but a dynamical variable which is constant 
only on-shell. This difference is vital, 
because it means that the quantum theory admits linear superpositions of 
states with different values of the gravitino mass and cosmological constant. 
 
The Ogievetsky-Sokatchev model also has two other important new features.
Firstly, it is invariant under the gauge transformation 
\begin{equation}                                          
M^m\mapsto M^m+\delta M^m,\ \ \ \ 
\delta M^m = e\epsilon^{mjkl} \partial_j A_{kl}
\label{eq:gauge_symmetry}
\end{equation} 
where $A_{kl}$ is an anti-symmetric tensor density. Secondly, owing 
to the elimination of the 
$\lambda,\lambda^*$ terms from the action, the Ogievetsky-Sokatchev model 
also has the global $U(1)$ symmetry 
\begin{equation} 
\psi\mapsto e^{i\phi}\psi, \ \ \ \ \ M^m\mapsto e^{2i\phi} M^m.
\label{eq:U1_symmetry} 
\end{equation}                                           
In the conventional formulation of supergravity this symmetry is broken 
for $\lambda\neq 0$.  
                                                  
We wish to determine whether the dynamical nature of $\Lambda$ 
in Ogievetsky-Sokatchev supergravity leads to a natural definition 
of cosmological time, as in unimodular general relativity. To this end 
we now consider the canonical formulation of the theory, focussing  
on those points which arise from the replacement of 
the scalars $M$, $M^*$ by the vector densities $M^m, M^{*m}$ as 
independent fields.  (See \cite{D'Eath} for a canonical description of the 
conventional theory.) 
 
We begin by making a 3+1 space-time split, with spacelike coordinates denoted 
by hatted latin indices $(\hat m,\hat n,\ldots)$ and the timelike 
coordinate denoted by the index $t$.                                    
For example, the tetrad $e^n{}_a$ is split into a timelike part 
$e^t{}_a = -n_a/N$ and a spatial part 
$e^{\hat n}{}_a = n_a (N^{\hat n} /N) 
+  h^{\hat n\hat m} e_{\hat m a}$ where $N$ is the lapse 
funtion, $N^{\hat m}$ is the shift vector, 
$h_{\hat m\hat n}= e_{\hat m}{}^a e_{n\hat a}$ is the spatial 3-metric, 
$h^{\hat m\hat n}$ is its inverse, and $n_a$ is defined so that 
$n_an^a= -1$ and $e_{\hat m}{}^a n_a=0$. 

Similarly, the vector densities $M^m,M^{*m}$ consist of spatial 
components $M^{\hat m}$, $M^{*\hat m}$ which are non-dynamical, and 
time-like components $M^t,M^{*t}$ which \it are \rm dynamical and 
have canonical momenta $p=-{1\over 3} M^*$, $p^* =  -{1\over 3}M$.  

Within any spatial hypersurface $\Sigma(t)$,  the spatially varying part 
of $M^t$ can be gauged away by a transformation of the form 
(\ref{eq:gauge_symmetry}). 
(One simply chooses the gauge transformation parameter as  
$A_{\hat k\hat l}= e^{-1} 
\epsilon_{t\hat k\hat l\hat m} B^{\hat m}$ where 
$2\partial_{\hat n}\partial_{\hat m} B^{\hat m} +\partial_{\hat n} M^t=0$.) 
After the removal of the gauge degrees of freedom, all that remains of $M^t$ 
is the spatially constant part whose integral is  
\begin{equation} 
q(t) \equiv  \int_{\Sigma(t)} M^t .
\end{equation}  

The Rarita-Schwinger fields are also split into timelike parts $\psi_t$, 
$\bar\psi_t$ and spatial parts 
$\psi_{\hat m}$, $\bar\psi_{\hat m}$. It is convenient to eliminate 
$\psi_t{}^\alpha, \bar\psi_{t\dot\alpha}$ in favour of the spinor
\begin{equation}  
\chi^\alpha   \equiv  \psi_t{}^\alpha 
-{2\over 3} N h^{\hat m\hat n} e_{\hat n a} n_b \psi_{\hat m}{}^{\beta} 
\sigma^{ab}{}_\beta{}^\alpha      
 - {1\over 3} N^{\hat m} \psi_{\hat m}{}^\alpha 
- {1\over 3} N^{\hat m} e_{\hat m c}h^{\hat \ell\hat n} e_{\hat n a } 
\psi_{\hat \ell}{}^\beta
\sigma^{ac}{}_\beta{}^\alpha \label{eq:new_lag_mult1} 
\end{equation} 
and its hermitian conjugate $\bar\chi_{\dot\alpha}$. This definition leads to 
the useful identity 
$e\psi_a\sigma^{ab}\psi_b = -2 h^{1\over 2} (\chi\sigma^{ab}\psi_{\hat m}) 
n_a h^{\hat m\hat n} e_{\hat n b}$, where $h\equiv \det [h_{\hat m\hat n}]$.

We also eliminate $N$ in favour of the variable $\widetilde N\equiv 
Nh^{1\over 2}$. The Hamiltonian is then
\begin{eqnarray}  
H & = &  \int d^3 x \Bigl\{   
M^{\hat m}\partial_{\hat m} p + M^{* \hat m}\partial_{\hat m} p^*  
+ \omega_t{}^{ab} J_{ab}  \nonumber\\  
& & + \widetilde N [ h^{-{1\over 2}}  {\cal H} - 3 p^* p 
+{2\over 3} h^{-{1\over 2}}h^{\hat m\hat n} e_{\hat n a} n_b ( 
\psi_{\hat m}{}^{\beta} \sigma^{ab}{}_\beta{}^\alpha S_\alpha       
+\bar\psi_{\hat m\dot\beta}  \bar\sigma^{ab}{}^{\dot\beta}{}_{\dot\alpha} 
S^{\dot \alpha})  ] \nonumber \\    
& & + N^{\hat m} [  {\cal H}_{\hat m}
+{1\over 3} (  \psi_{\hat m}{}^\alpha  + e_{\hat m c}h^{\hat \ell\hat n} 
e_{\hat n a } \psi_{\hat \ell}{}^\beta  \sigma^{ac}{}_\beta{}^\alpha ) 
S_\alpha + {1\over 3} ( \bar\psi_{\hat m\dot\alpha} 
+ e_{\hat m c}h^{\hat\ell\hat n} e_{\hat n a } 
\bar\psi_{\hat\ell\dot\beta} \bar\sigma^{ac}{}^{\dot\beta}{}_{\dot\alpha}) 
\bar S^{\dot\alpha} ] \nonumber\\
& &  + \chi^\alpha  ( S_\alpha 
- 2 p h^{1\over 2}  n_a h^{\hat m \hat n} e_{\hat n b}
\sigma^{ab}{}_\alpha{}^\beta  \psi_{\hat m\beta}) \nonumber \\
& & + \bar\chi_{\dot\alpha} ( \bar S^{\dot \alpha} 
- 2 p^* h^{1\over 2} n_a h^{\hat m \hat n} e_{\hat n b}
\bar\sigma^{ab}{}^{\dot\alpha}{}_{\dot\beta} 
\bar\psi_{\hat m}{}^{\dot\beta} ) \Bigr\}  
\label{eq:Hamiltonian} 
\end{eqnarray} 
where the expressions for the quantities 
${\cal H}(x)$, ${\cal H}_{\hat m}(x)$, $S_{\alpha}(x)$, 
$\bar S^{\dot\alpha}(x)$, and $J_{ab}(x)$ 
are familiar from the standard formulation of canonical supergravity 
\cite{D'Eath}, in which they play the roles of Hamiltonian, momentum, 
supersymmetry and Lorentz constraints respectively. 

The momenta conjugate to $\widetilde N(x)$, $N^{\hat m}(x)$, $\chi^\alpha(x)$, 
$\bar\chi_{\dot\alpha}(x)$, 
$\omega_t{}^{ab}(x)$, $M^{\hat m}(x)$ and $M^{*\hat m}(x)$ are found to vanish, 
indicating that these variables act as Lagrange multipliers in 
(\ref{eq:Hamiltonian}).  The requirement that $H$ be stationary with respect 
to these Lagrange multipliers then leads to a set of secondary constraints at 
each point $x$. In particular, $M^{\hat m}$ and $M^{*\hat m}$ enforce the 
constraints 
$\partial_{\hat m} p = \partial_{\hat m} p^* = 0 $ 
indicating that $p,p^*$ are spatially constant. In fact the parameters $p(t),
p^*(t)$ are simply the momenta conjugate to the canonical coordinates 
$q(t), q^*(t)$.  

The quantity $\Lambda = -3p^*p$ has physical significance, as it 
plays the role of the cosmological constant in the field equations 
derived from (\ref{eq:Lagrangian}). We therefore eliminate $p(t),p^*(t)$ 
in favour of real dynamical variables $\Lambda(t)$ and $\theta(t)$, defined 
so that $p= ( - \Lambda/ 3)^{1\over 2} e^{i\theta}$. The equations of 
motion then ensure that $\Lambda$ and $\theta$ remain constant along 
classical trajectories.

In unimodular general relativity, $\Lambda$ is canonically 
conjugate to a variable which can be interpreted as the cosmological time 
parameter. In the present case, the variables $\Lambda(t)$ and $\theta(t)$ 
can be identified respectively as the momenta $\Pi_T$, $\Pi_Q$ conjugate to 
the new canonical coordinates 
\begin{eqnarray}  
T(t) & = &  -{1\over 6p^*p} ( pq  + p^*q^*)  \\
Q(t) & = &   i(pq -p^*q^*) 
\end{eqnarray}                                               
as can be seen by calculating the Dirac bracket relations. 

The analogy with unimodular general relativity raises the hope that the new 
variable $T$ might play the role a cosmological time parameter.
This hope is fulfilled. The equations of motion derived from 
(\ref{eq:Lagrangian}) imply that 
\begin{equation} 
{ dT\over dt} = \int\limits_{\Sigma(t)} d^3x \, 
\left[  e -{ h^{1\over 2}\over\Lambda} ( p \chi\sigma^{ab}\psi_{\hat m} + 
p^*\bar\chi\bar\sigma^{ab}\bar\psi_{\hat m} ) n_a h^{\hat m\hat n} 
e_{\hat n b} \right]. 
\end{equation}                                                                          
Integrating and making use of the supersymmetry constaints (enforced 
in (\ref{eq:Hamiltonian}) by the Lagrange multipliers 
$\chi^\alpha,\bar\chi_{\dot\alpha}$) 
one obtains the weak equalities \cite{Dirac}   
\begin{equation} 
T(t) \approx\int_{\cal M} d^4 x \left[ e - 
{1\over 2\Lambda}( \chi^\alpha S_\alpha 
+ \bar\chi_{\dot\alpha} \bar S^{\dot\alpha})  \right] 
\end{equation}                               
where $\cal M$ denotes the 4-volume between the hypersurfaces 
$\Sigma(t_0)$ and $\Sigma(t)$, with $t_0$ defined so that $T(t_0) =0$.  
Similarly, one finds that 
\begin{equation} 
Q(t)-Q(t_0) \approx -i \int_{\cal M} d^4 x\, 
(\chi^\alpha S_\alpha -\bar\chi_{\dot\alpha}\bar S^{\dot\alpha}) . 
\end{equation}        

In order to extract the gauge-invariant content of $T(t)$ and $Q(t)$, we may
now choose a gauge with $\chi^\alpha=\bar\chi_{\dot\alpha}= 0$ everywhere. The 
equations of motion then imply that $Q(t)$ is constant while $T(t)$ is the 
invariant 4-volume of spacetime preceeding the hypersurface $\Sigma(t)$. In 
this gauge, therefore, $T(t)$ coincides classically with the cosmological time 
parameter that arises both in unimodular general relativity \cite{Unruh, 
Henneaux,UnruhWald} and in Sorkin's sum-over-histories approach \cite{Sorkin}. 
Note that $T$ is a monotonically increasing function along any classical 
trajectory and so can indeed be used to parametrise this trajectory.
                     
As well as imposing gauge conditions on $\chi^\alpha$ $\bar\chi_{\dot\alpha}$, 
it is also necessary to fix the other Lagrange multipliers $\widetilde N(x)$, 
$N^{\hat m}(x)$ and $\omega_t{}^{ab}(x)$ at each spacetime point $x$ in order 
that the equations of motion have a unique solution and the classical 
evolution is well-defined. (It is immaterial \it how \rm these Lagrange 
multipliers are chosen, provided that $\widetilde N>0$.)  Then through each 
point in phase space at which the constraints hold, 
there passes a unique classical trajectory parametrised by $T$.
The closure of the Dirac constraint algebra ensures that the whole trajectory 
will be confined to the region in phase space where the classical constraints 
are satisfied. 
                                        
Having removed the ambiguities from the classical evolution equations by 
specifying the Lagrange multipliers, we now proceed to the quantum theory. 
The constraints become conditions on the wave function $\Psi$, their precise 
form depending on the representation used. An  
obvious choice is the $(\Lambda,\theta)$ representation, with a wave 
function $\Psi(\Lambda,\theta;e_{\hat m}{}^a,
\bar\psi_{\hat m}{}^{\dot\alpha}]$.
However, more insight into the nature of the supersymmetry constraints is 
obtained by switching to the $(\Lambda,Q)$ representation, with 
$\Psi(\Lambda,\theta;e_{\hat m}{}^a,\bar\psi_{\hat m}{}^{\dot\alpha}]$ 
decomposed as a linear combination of eigenmodes $\Psi_N$ of the 
operator $ Q= i\hbar\partial/\partial\theta$: 
\begin{equation}                                                
\Psi(\Lambda,\theta;e_{\hat m}{}^a,\bar\psi_{\hat m}{}^{\dot\alpha}] 
= \sum_{N=-\infty}^\infty 
e^{-iN\theta}\,  \Psi_N(\Lambda;e_{\hat m}{}^a,
\bar\psi_{\hat m}{}^{\dot\alpha}].
\end{equation}  
                                    
At each point $x$ there is a family of constraints, which in this 
representation have the form 
\begin{eqnarray}
h^{-{1\over 2}}{\cal H}\Psi_N & = &   
- \left[ {2\over 3} h^{-{1\over 2}}h^{\hat m\hat n} e_{\hat n a} n_b ( 
\psi_{\hat m}{}^{\beta} \sigma^{ab}{}_\beta{}^\alpha S_\alpha       
+\bar\psi_{\hat m\dot\beta}  \bar\sigma^{ab}{}^{\dot\beta}{}_{\dot\alpha} 
\bar S^{\dot \alpha}) + \Lambda\right] \Psi_N  \label{eq:Ham_constraint}\\   
{\cal H}_{\hat m} \Psi_N & = & 
-{1\over 3} \Bigl[  ( \psi_{\hat m}{}^\alpha  
+ e_{\hat m c}h^{\hat \ell\hat n} e_{\hat n a } 
\psi_{\hat \ell}{}^\beta  \sigma^{ac}{}_\beta{}^\alpha ) S_\alpha \nonumber \\
& & \ \ \ \ \ \ \ \ \ 
 + ( \bar\psi_{\hat m\dot\alpha} + e_{\hat m c}h^{\hat\ell\hat n} e_{\hat n a } 
\bar\psi_{\hat\ell\dot\beta} \bar\sigma^{ac}{}^{\dot\beta}{}_{\dot\alpha} ) 
\bar S^{\dot\alpha} \Bigr] \Psi_N  \label{eq:mom_constraint}\\  
S_\alpha\Psi_N &= &   2 \left( - {\Lambda\over 3} \right)^{1\over 2} 
 h^{1\over 2}  n_a h^{\hat m \hat n} e_{\hat n b}
\sigma^{ab}{}_\alpha{}^\beta  \psi_{\hat m\beta} \Psi_{N+1}  
\label{eq:susy_constraint1}\\     
\bar S^{\dot \alpha}\Psi_N &= &  2 \left( - {\Lambda\over 3} 
\right)^{1\over 2} h^{1\over 2} n_a h^{\hat m \hat n} e_{\hat n b}
\bar\sigma^{ab}{}^{\dot\alpha}{}_{\dot\beta} \bar\psi_{\hat m}{}^{\dot\beta}
\Psi_{N-1} \label{eq:susy_constraint2}\\  
J_{ab}\Psi_N & =  & 0. \label{eq:AM_constraint} 
\end{eqnarray}  
(For brevity, we have suppressed the spatial dependence of the operators 
$h^{-{1\over 2}}(x)$, ${\cal H}(x)$,  ${\cal H}_{\hat m}(x)$, $J_{ab}(x)$, 
$S_\alpha(x)$, $\bar S^{\dot\alpha}(x)$, $h^{mn}(x)$, $e_{\hat n a}(x)$, 
$n_a(x)$, $\psi_{\hat m}{}^{\alpha}(x)$ and $\bar\psi_{\hat m 
\dot\alpha}(x)$.)      
The argument $\Lambda$ is restricted to the non-positive part of 
the real axis, and each $\Psi_N$ is required to vanish at $\Lambda=0$. This 
boundary condition can be derived from the continuity of the wave function 
in the $(p,p^*)$ representation, and ensures that the operator 
$T= i\hbar\partial/\partial\Lambda$ is self-adjoint.

An alternative description of the quantum theory can be obtained using 
the $(T,Q)$ representation, in which the wave function is defined by the 
Fourier transform
\begin{equation} 
\Psi_N(T;e_{\hat m}{}^a,\bar\psi_{\hat m}{}^{\dot\alpha} ] 
\equiv {1\over \sqrt{2\pi}} \int_{-\infty}^0 d\Lambda \, \exp 
\left\{ {i\over \hbar} T \Lambda \right\}\,   
\Psi_N(\Lambda;e_{\hat m}{}^a,\bar\psi_{\hat m}{}^{\dot\alpha} ]                
\end{equation}                                                      
and $\Lambda$ is represented by the operator $-i\hbar \partial/\partial T$. 
In this representation, the Hamiltonian constraints (\ref{eq:Ham_constraint}) 
take the form of a family of Schr\"odinger equations (one at each point $x$) 
describing the evolution of the cosmological wave function with respect to the 
cosmological time parameter $T$: 
\begin{equation} 
i\hbar {\partial \Psi_N\over \partial T}  =  
h^{-{1\over 2}} \left\{ {\cal H} + {2\over 3} h^{\hat m\hat n}
 e_{\hat n a} n_b \Bigl[  
\psi_{\hat m}{}^{\beta} \sigma^{ab}{}_\beta{}^\alpha S_\alpha       
+\bar\psi_{\hat m\dot\beta}  \bar\sigma^{ab}{}^{\dot\beta}{}_{\dot\alpha} 
\bar S^{\dot \alpha}\Bigr]\,  \right\}  \Psi_N .
\label{eq:Schrodinger} \\
\end{equation} 
                                                   
The momentum and angular momentum constraints have the same form 
(\ref{eq:mom_constraint},\ref{eq:AM_constraint}) in the 
$(T,Q)$ representation as in the $(\Lambda,Q)$ representation. However, the 
supersymmetry constraints are awkward  to express in the 
$(T,Q)$ representation, as they involve the square root of the operator 
$\Lambda = -i\hbar \partial/\partial T$; when considering these constraints, 
it is convenient to return to the $(\Lambda,Q)$ representation.  

Assuming that the operator-ordering is chosen so that 
$\cal H$ is self-adjoint and $S^\alpha, \bar S_{\dot\alpha}$ are mutually 
adjoint with respect to the measure on the configuration space 
$(e_{\hat m}{}^a,\bar\psi_{\hat m}{}^{\dot\alpha})$, then it follows from 
the Schr\"odinger equations (\ref{eq:Schrodinger}) that the integral of the 
quantity $\Psi_N^*\Psi_N$ is conserved with respect to $T$. 
Moreover, once we have integrated over the fermionic degrees of freedom and 
summed over all values of $N$, this quantity is real and non-negative, and so 
is naturally interpreted as the probability density function for the Universe.  

The momentum constraints (\ref{eq:mom_constraint}) imply that $\Psi$ depends 
only on $T$ and the equivalence class of configurations related to  
$(e_{\hat m}{}^a,\bar\psi_{\hat m}{}^{\dot\alpha})$ 
by spatial diffeomorphisms.
An argument by Kucha\v r \cite{Kuchar2} can be adapted to show that these 
equivalence classes are not Dirac observables. However, this is not an obstacle 
to the interpretation of the wave function suggested above; in a parametrised 
theory, the wave function arguments need not be observables. (For example, 
see \cite{Marolf}.) 

This is illustrated by the parametrised description of particle dynamics 
in a one-dimensional potential. In this description both $t(\tau)$ and 
$x(\tau)$ are coordinates, with conjugate momenta $\pi(\tau), p(\tau)$. 
The Hamiltonian 
\begin{equation}
H = N \bigl[ \pi + {1\over 2}p^2 + V(x) \bigr]. 
\end{equation}        
contains a Lagrange multiplier $N$ enforcing the constraint 
$\phi \equiv  \pi + {1\over 2}p^2 + V(x) \approx 0$     
which in the quantum theory imposes the condition 
\begin{equation} 
-\pi\Psi= \bigl[ {p^2\over 2}  + V(x) \bigr] \Psi. 
\label{eq:Schro} 
\end{equation}                              
In the $(t,x)$ representation, the momenta are given by the operators 
$\pi=-i \hbar\partial/\partial t$ and $p= -i \hbar\partial/\partial x$
and so (\ref{eq:Schro}) is just the Schr\"odinger equation. We must
therefore adopt the conventional interpretation of $\Psi(t,x)$ as the amplitude 
at time $t$ for observing the particle at position $x$. 

It is easily seen that the variables $t$ and $x$ have non-vanishing 
Dirac brackets with the constraint $\phi$, and so are not Dirac observables. 
However this fact does not
prevent us from adopting the conventional interpretation of the wave function 
$\Psi(t,x)$ ; it simply reflects the breaking of time-translation symmetry by 
the act of measurement at a definite instant. 

Similarly, in the present theory, the non-commutation of the wave function 
arguments with the Hamiltonian and supersymmetry constraints 
reflects the fact that measurements are to be made on a definite hypersurface 
$\Sigma$ and in a definite supersymmetry gauge. Once again, this does not 
prevent us from adopting the conventional interpretation of the wave function.

It should be noted here that the dynamical variable $T$ need not be assumed to 
play any special role in the identification of the hypersurface $\Sigma$ in the 
classical theory. (For example, one might specify the embedding of the 
hypersurface 
by giving the spacetime coordinates of each point on $\Sigma$.) However, since 
$T$ increases monotically along classical trajectories (at least in the gauge 
$\chi=\bar \chi= 0$), there is a strong temptation to view it as a Heraclitean 
time parameter labelling the hypersurfaces in some foliation of 
spacetime \cite{UnruhWald}.

In the context of unimodular gravity, it has been argued that the specification
of the cosmological time $T$ is insufficient to identify the hypersurface
on which the measurements are to be made \cite{Kuchar1,Kuchar2}. While this is 
certainly true, the argument no longer holds if the Lagrange multipliers 
$\widetilde N(x)$ and $N^{\hat m}(x)$ are fixed in advance, as in the present 
approach; then the classical evolution is completely determined and each 
value of $T$ specifies a unique hypersurface. Thus, one can identify a 
particular hypersurface by specifying a value of the parameter $T$ \it 
and \rm the Lagrange multipliers at each spacetime point. 
                                      
It is clear that the choice of Lagrange multipliers makes no difference at all 
to the quantum constraints (\ref{eq:Ham_constraint}-\ref{eq:AM_constraint}),
or to the evolution of the wave function $\Psi$ with respect to the 
cosmological time parameter $T$. Hence the the transition amplitude between 
two specified hypersurfaces is independent of any coordinate conditions which 
may be imposed on the interpolating spacetimes. The quantum theory therefore 
escapes the ``multiple choice problem", which arises in most other 
approaches to quantum gravity in which time is found among the canonical 
variables \cite{Kuchar1}.   

In conclusion, we have shown that the canonical quantisation of 
Ogievetsky-Sokatchev supergravity leads directly to a time-dependent wave 
function with a straightforward probabilistic interpretation. 
It is natural to ask whether this result applies only in $N=1$ supergravity, 
or is enjoyed by a wider class of supersymmetric theories. The essential 
ingredient 
appears to be the replacement of an auxiliary scalar field by the divergence 
of a vector field with well-defined supersymmetric transformation properties.
This is not a very stringent requirement, and can probably be satisfied by 
a wide variety of locally supersymmetric theories. Any such theory will
have features similar to those outlined above. 
 
\centerline {\bf Acknowledgements}
This work was supported by an Australian Research Council Institutional Grant, 
and by the Deutsche Forschungsgemeinschaft through the Sonderforschungsbereich 
237 ``Unordnung und Grosse Fluktuationen".

\vfill
\end{document}